\documentclass[twocolumn,showpacs,preprintnumbers,amsmath,amssymb,aps,prc,superscriptaddress,reprint,floatfix]{revtex4-2}
\usepackage{graphicx}
\usepackage{dcolumn}
\usepackage{bm}
\usepackage[utf8]{inputenc}
\usepackage{multirow}
\usepackage{tabularx}
\usepackage{microtype}
\microtypesetup{protrusion=true,expansion=true,tracking=true,kerning=true}
\microtypecontext{spacing=nonfrench}
\usepackage{hyperref}
\hypersetup{unicode=false,colorlinks=true,breaklinks=true,citecolor=blue,urlcolor=blue,linkcolor=blue}
\begin{document}

\title{Monte Carlo rate uncertainty of the $^{8}$Li(n,$\gamma$)$^{9}$Li reaction within $R$-matrix framework}

\author{Sk Mustak Ali}
\email{mustak.ali@saha.ac.in}
\thanks{(Corresponding author)}
\affiliation{Saha Institute of Nuclear Physics, 1/AF Bidhannagar, Kolkata 700064, India}

\author{Rajkumar~Santra}
\affiliation{Department of Physics, University of Calcutta, Kolkata-700009, India}

\date{\today}

\begin{abstract}
The $^{8}$Li$(n,\gamma)^{9}$Li reaction is considered
significant for the synthesis of nuclei beyond the $A=8$ stability gap in
inhomogeneous big-bang nucleosynthesis models, as well as in $r$-process
nucleosynthesis scenarios. However, direct
measurement of this reaction is precluded by the short half-life of $^{8}$Li and
the absence of a neutron target. Consequently, existing reaction rate estimates based on
indirect experimental methods and theoretical calculations differ by orders of
magnitude. In the present work, the $^{8}$Li(n,$\gamma$)$^{9}$Li neutron-capture cross section and the
corresponding thermonuclear reaction rate are evaluated within a phenomenological
$R$-matrix framework, including both non-resonant direct capture (DC) and
resonant capture through the $5/2^{-}$ state at $E_{x} = 4.30$~MeV. The uncertainties associated with the $R$-matrix input
parameters are propagated using Monte Carlo sampling, while the sensitivity to
the channel radius is treated as an $R$-matrix model uncertainty.
The effective total uncertainties in the calculated cross sections and reaction
rates are obtained by adding these two uncertainty contributions in quadrature.
We obtain a
total capture rate of
$983.9^{+410.9}_{-261.5}~\mathrm{cm^{3}\,mol^{-1}\,s^{-1}}$ at $T = 1$~GK, with
DC dominating at low temperatures ($T=0.01-0.4$~GK) and the $5/2^{-}$ resonance at higher temperatures ($T=0.5-5$~GK). The present results are consistent with the upper limit of Kobayashi et al.~\cite{Kobayashi2003}, which previous theoretical predictions exceed by factors of 3-50.
\end{abstract}

\maketitle

\section{Introduction}
The radiative capture reaction $^{8}$Li(n,$\gamma$)$^{9}$Li has gained significant attention due to its astrophysical implications. In particular, this reaction is important in the nucleosynthesis of light elements hindered by the stability gap at mass number $A=8$.
In the context of the inhomogeneous big-bang nucleosynthesis (IBBN) scenario, where the early universe is characterized by regions of neutron-rich and proton-rich environments, the $^{8}$Li(n,$\gamma$)$^{9}$Li reaction is a crucial component. In neutron-rich pockets, these inhomogeneous models predict relatively higher abundances of nuclides with $A>8$, contrary to the standard BBN model, via the $^7$Li(n,$\gamma$)$^8$Li($\alpha$,n)$^{11}$B(n,$\gamma$)$^{12}$B($\beta^-$)$^{12}$C chain~\cite{Malaney1989, Kajino1990, Mao1991, Boyd1992, Rauscher1994}. However, the neutron-capture reaction $^{8}$Li(n,$\gamma$)$^{9}$Li can greatly reduce the production of heavier nuclei in the above chain by competing with the $^{8}$Li($\alpha$,n)$^{11}$B channel.
Further, $^{8}$Li(n,$\gamma$)$^{9}$Li is a vital link in the $^{4}$He(2n,$\gamma$)$^6$He(2n,$\gamma$)$^8$He($\beta^-$)$^{8}$Li(n,$\gamma$)$^{9}$Li$(\beta^-)$$^9$Be pathway, offering an alternative route to the 3-$\alpha$ process in the $r$-process for type-II supernovae~\cite{Gorres1995}. 

Direct experimental measurement of this reaction is currently unfeasible due to the absence of a neutron target and the short half-life of $^8$Li, $T_{1/2}\approx 840$~ms. Moreover, the low reaction cross section at astrophysically relevant energies makes such a direct measurement extremely challenging. Hence, indirect experimental techniques such as the asymptotic normalization coefficient (ANC) method and the Coulomb dissociation approach have been used to study this reaction.
Upper limits of the reaction rates were determined through two Coulomb dissociation experiments by Zecher~\textit{et al.}~\cite{Zecher1998} and Kobayashi~\textit{et al.}~\cite{Kobayashi2003}, employing a $^{9}$Li beam traversing the virtual photon field of high-$Z$ targets. However, these upper limits differed by an order of magnitude.
Additionally, transfer-reaction measurements such as $^{8}$Li(d,p)$^{9}$Li~\cite{Li2005} and $^{9}$Be($^8$Li,$^9$Li)$^{8}$Be~\cite{Guimaraes2007} were carried out to extract the neutron spectroscopic factors of $^9$Li. These spectroscopic factors were then used to calculate the neutron-capture cross sections in the potential model~\cite{Li2005,Guimaraes2007}. The rates obtained in these transfer studies were consistent with the upper limits reported by Zecher~\textit{et al.}~\cite{Zecher1998}, but were $4-5$ times higher than the upper limit reported by Kobayashi~\textit{et al.}~\cite{Kobayashi2003}.

Theoretical predictions for the $^{8}$Li(n,$\gamma$)$^{9}$Li reaction exhibit substantial discrepancies across different nuclear models. Neutron-capture rates derived from systematics based on other nuclei~\cite{Malaney1989,Rauscher1994}, as well as calculations using the shell model~\cite{Mao1991,Ma2012}, microscopic cluster model~\cite{Descouvemont1993}, and potential models~\cite{Bertulani1999,Mohr2003, Banerjee2008, Dubovichenko2016}, span orders of magnitude, from $\sim2\times 10^3$ to $4\times 10^4$~cm$^3$mol$^{-1}$s$^{-1}$ at temperature $T_9 = 1$ ($T_9\equiv T/10^9$~K). A recent \textit{ab initio} no-core shell model with continuum (NCSMC) calculation~\cite{McCracken2021} predicts a significantly higher $^{8}$Li(n,$\gamma$)$^{9}$Li reaction rate compared to those obtained from potential models and Gamow shell model calculations~\cite{Dong2022,Dong2023Erratum}.
A summary of the $^{8}$Li(n,$\gamma$)$^{9}$Li reaction rates from various theoretical and experimental studies is presented in Table~\ref{tab:9Li rates}.

The significant discrepancies in the $^{8}$Li(n,$\gamma$)$^{9}$Li cross sections reported by various theoretical models and indirect experimental measurements motivate a reevaluation of this reaction within the $R$-matrix framework. In this work, we perform an $R$-matrix analysis of the $^8$Li(n,$\gamma$)$^9$Li reaction, incorporating available spectroscopic information for the bound and resonant states in $^9$Li from the literature. Both direct and resonant capture components are studied over a broad range of center-of-mass energies, $E_\text{c.m.}$, and the corresponding reaction rates are calculated for astrophysically relevant temperatures ($T_9=0.01-5$). To quantify the uncertainties in the calculated cross sections and reaction rates, we perform Monte Carlo sampling of the relevant $R$-matrix input parameters, including the excited state energies ($E_x$), neutron ($\Gamma_n$) and gamma widths ($\Gamma_\gamma$), and asymptotic normalization coefficients ($C$), within their reported uncertainties. 
In addition to these parameter uncertainties, we estimate the model uncertainty associated with the $R$-matrix description, arising primarily from the sensitivity of the calculated results to reasonable variations of the \textit{channel radius}. The total uncertainty is then obtained by combining this model uncertainty with the Monte Carlo propagated parameter uncertainty.
The resulting cross sections and reaction rates are compared with existing experimental and theoretical studies.
\begin{table*}[t]
\caption{\label{tab:9Li rates} The $^8$Li(n,$\gamma$)$^9$Li reaction rate at $T_9 = 1.0$ ($T_9\equiv T/10^9$~K) from previous experimental and theoretical studies.}
\begin{ruledtabular}
\begin{tabular}{lcccc}
Reference & Year & Rate (cm$^3$\,mol$^{-1}$\,s$^{-1}$) & Experiment & Theory \\
\hline
Zecher~\textit{et al.}~\cite{Zecher1998}    & 1998 & $<$7200         & Coulomb dissociation &        \\
Kobayashi~\textit{et al.}~\cite{Kobayashi2003} & 2003 & $<$790          & Coulomb dissociation &        \\
Li~\textit{et~al.}~\cite{Li2005} & 2005 & 4000 & Transfer reaction&\\
Guimaraes~\textit{et~al.}~\cite{Guimaraes2007} & 2007 & 3270 & Transfer reaction &\\
\hline
Malaney and Fowler~\cite{Malaney1989}       & 1989 & 43000         &       &\textsuperscript{\textdagger}  \\
Mao and Champagne~\cite{Mao1991}            & 1991 & 27778\textsuperscript{a}, 21042\textsuperscript{b} &       & Shell model \\
Descouvemont~\cite{Descouvemont1993}        & 1993 & 5280            &       & Microscopic cluster model \\
Rauscher~\textit{et al.}~\cite{Rauscher1994} & 1994 & 4500            &       & \textsuperscript{\textdagger}\\
Bertulani~\cite{Bertulani1999}              & 1999 & 2200            &       & Potential model \\
Banerjee~\textit{et al.}\cite{Banerjee2008} & 2008 & 2900 & & FRDWBA\textsuperscript{$\ast$}\\
Ma~\textit{et~al.}~\cite{Ma2012} & 2012 & $<4300$\textsuperscript{c} & & Shell model\\
Dubovichenko~\textit{et al.}~\cite{Dubovichenko2016} & 2016 & 5900 & & Modified potential cluster model\\
McCracken~\textit{et~al.}~\cite{McCracken2021} & 2021 & 11770 & & \textit{ab initio} NCSMC\textsuperscript{$\ddagger$} \\
Dong~\textit{et~al.}~\cite{Dong2022,Dong2023Erratum} & 2022/23 & $\sim 5200$ & & GSM-CC\textsuperscript{$\S$} \\
\end{tabular}
\end{ruledtabular}
\begin{flushleft}
\textsuperscript{\textdagger}based on the information existing in other nuclei.\\
\textsuperscript{$\ddagger$} No-core shell model with continuum.\\
\textsuperscript{$\ast$} finite range distorted wave Born approximation.\\
\textsuperscript{$\S$}Gamow shell model (GSM) in the coupled-channel representation.\\
\textsuperscript{a} $p$ shell, \textsuperscript{b} $spd$ shell, \textsuperscript{c} total neutron capture rate.
\end{flushleft}
\end{table*}

The paper is organized as follows. Section~\ref{sec:formalism}
describes the reaction-rate formalism and the phenomenological
$R$-matrix framework used in the present calculation. The Monte Carlo
sampling procedure and the adopted probability distributions for the
$R$-matrix input parameters are discussed in
section~\ref{sec:MC_uncertainty}. Section~\ref{sec:model_uncertainty}
presents the sensitivity of the calculated cross sections on channel radius and discusses the associated $R$-matrix
model uncertainty. The calculated cross sections, thermonuclear
reaction rates, and parameter-sensitivity results are presented and
compared with previous experimental and theoretical studies in
section~\ref{sec:Results}. The conclusions are summarized
in section~\ref{sec:Conclusion}.

\section{REACTION RATE FORMALISM} \label{sec:formalism}

\subsection{R-Matrix Framework}\label{sec:R-matrix}
Phenomenological $R$-matrix analysis has been carried out for the $^8$Li(n,$\gamma$)$^9$Li reaction over the center-of-mass energy range $E_\text{c.m.} = 0.01–1.32$ MeV to evaluate reaction rates at astrophysically relevant temperatures. The present calculations include both non-resonant ``direct capture'' (DC) and resonant contributions. In this context, direct capture refers to the transition from an initial state to a final state in a non-resonant manner, without the formation of an intermediate compound-nuclear resonance~\cite{deBoer2017}.

The analysis employs the \texttt{AZURE2} code~\cite{AZURE}, which implements the multilevel, multichannel $R$-matrix formalism originally developed by Lane and Thomas~\cite{Lane1958}.
A detailed overview of the $R$-matrix approach is provided in Ref.~\cite{deBoer2017}.
The $R$-matrix formalism divides the nuclear configuration space into two distinct regions: an internal region, where nuclear interactions dominate, and an external region, where the long-range Coulomb interaction is dominant. These regions are separated by a channel radius ($r_c$). 
Accordingly, the radiative capture cross section is divided
into an internal capture component ($r < r_c$) and an external capture component ($r > r_c$).
The non-resonant DC component of the present calculation is given by the
external capture (EC) term in the hard-sphere formulation, with its
magnitude governed by the ANC of the final bound state~\cite{AZURE}.  In fitted
$R$-matrix analyses, the internal non-resonant contribution is simulated by high-energy background poles constrained by
data. As the present calculation involves no
fitting due to the unavailability of precise experimental data, such background poles are not employed, and the DC cross section
corresponds to the EC alone.

In the present analysis, the sensitivity of the external-capture (EC) component to the channel radius, $r_c$, is investigated over the range $r_c=4.0$--$5.0$~fm, as discussed in section~\ref{sec:model_uncertainty}. The channel radius is commonly chosen using the prescription
$r_c=1.4(A_1^{1/3}+A_2^{1/3})$, for the $^8$Li$+n$ system, this gives $r_c = 4.2$~fm. For comparison, previous $R$-matrix analysis of the neighboring $^7$Li$+n$ system adopted $r_c=4.3$~fm~\cite{Knox1981}. It should be noted that the channel radius is treated as a model parameter in the $R$-matrix framework. The variation of the EC cross section with $r_c$ is used to estimate the model uncertainty associated with the choice of channel radius, following the approach of Ref.~\cite{deBoer2014}.
There are eight input parameters sampled in the present $R$-matrix analysis, the level energies of the $^9$Li bound state at $E_x={2.69}$~MeV and the $5/2^-$ resonance at $E_x={4.30}$~MeV,
the ANCs ($C$) corresponding to the $3/2^-$ ground state (g.s.) and the $1/2^-$ excited state at $E_x = 2.69$ MeV, together with the partial neutron ($\Gamma_n$) and the $\gamma$-decay ($\Gamma_\gamma$) widths for the $M1$ and $E2$ transitions to g.s. and the $E2$ transition to the $E_x=2.69$~MeV state.  The Monte Carlo sampling of the input parameters is carried out at the central value of the $r_c$ range i.e.,  $r_c=4.5$~fm.

\subsection{Monte Carlo reaction rate}

Traditionally, uncertainties in $R$-matrix calculations are explored by varying 
individual input parameters within assumed ranges to generate ``upper'' and ``lower'' 
reaction-rate limits. However, such a procedure is not statistically rigorous as 
pointed out by Longland~\textit{et al.}~\cite{Longland2010}. Furthermore, the data 
reported by Zecher~\textit{et al.}~\cite{Zecher1998} are averaged over broad 
center-of-mass energy intervals of $E_\text{c.m.}=0-0.5$~MeV and 0.5--1.0~MeV, 
with only upper limits provided for the capture cross sections. Therefore, in the 
absence of reliable experimental cross-section data for $^8$Li(n,$\gamma$)$^9$Li, 
a direct Bayesian inference of the model parameters is not feasible. Instead, each 
$R$-matrix input parameter is assigned a probability density function (PDF) informed 
by experimental and theoretical constraints, and Monte Carlo (MC) sampling from these 
PDFs allows the uncertainties to be propagated consistently through the $R$-matrix 
formalism, yielding statistically meaningful uncertainty intervals for the predicted 
capture cross section and reaction rate.

The choice of sampling distributions follows the MC
reaction-rate prescription of Longland et al.~\cite{Longland2010}, where each nuclear-physics input quantity is assigned a
PDF appropriate to its physical origin and
allowed range. Quantities such as level energies are sampled from
Gaussian distributions because their uncertainties arise from the sum
of several independent experimental contributions, such as energy
calibration, yield determination, and reaction $Q$ values. A Gaussian
distribution is also adopted here for the ANCs, since the available
literature values are treated as a set of normally distributed estimates
and the mean and standard deviation of these values are used to
represent their dispersion.
In contrast, the partial widths $\Gamma_n$ and $\Gamma_\gamma$ are
strictly positive quantities. Thus, they are described by log-normal
distributions, since widths or resonance strengths are commonly derived
from products and ratios of positive quantities and a Gaussian sampling
could otherwise produce unphysical negative values. The log-normal
form also allows for asymmetric uncertainties.

In the present work, the MC approach is implemented through the python package
\texttt{BRICK}~\cite{BRICK} interfaced with the
\texttt{AZURE2}~\cite{AZURE} $R$-matrix code. Each input parameter
$\theta_i \in \{E_x,\, \Gamma_n,\, \Gamma_\gamma,\, C\}$ is treated as an independent random
variable drawn from an appropriate PDF reflecting the type and magnitude of its
uncertainty.
A total of $10000$ MC samples were generated by independently 
sampling each input parameter from its respective probability distribution, at $r_c=4.5$~fm. Each 
sample corresponds to a unique set of $R$-matrix input parameters, which was 
used to perform a complete $R$-matrix calculation with \texttt{AZURE2}, 
yielding corresponding cross sections $\sigma(E)$ and thermonuclear reaction 
rates $N_A\langle \sigma v \rangle$ over a temperature range of 
$T_9 = 0.01-5.0$. The reaction rate uncertainty bands are obtained from the ensemble of
10000 calculations by computing the 16\%, 50\%, and 84\% quantiles at
each temperature. The median value is defined by the 50\% quantile,
while the lower and upper MC uncertainties are defined by the 16\%
and 84\% quantiles, respectively. The details of the sampling distributions are discussed in section~\ref{sec:MC_uncertainty}.

\section{Monte Carlo uncertainty}\label{sec:MC_uncertainty}

The $^8$Li(n,$\gamma$)$^9$Li reaction can proceed through both non-resonant 
direct capture (DC) to the bound states in $^9$Li as well as resonant capture 
via the $5/2^-$ resonance at $E_x=4.30$~MeV. Literature values of the bound-state ANCs range 
from 0.96 to 1.23~fm$^{-1/2}$ for the $3/2^-$ ground state (g.s.) and from 0.298 to 
0.40~fm$^{-1/2}$ for the $1/2^-$ excited state at $E_x=2.69$~MeV (see 
Table~\ref{tab:9Li_parameters}). The sampling distributions for the ANCs are 
constructed as Gaussian distributions $\mathcal{N}(\mu,\sigma)$, where $\mu$ 
and $\sigma$ are the mean and standard deviation of the unweighted literature 
values listed in Table~\ref{tab:9Li_parameters}, ensuring that the full 
dispersion of reported values is reflected in the sampling distribution. 
Similarly, the level energies $E_{2.69}$ and $E_{4.30}$ are sampled from 
Gaussian distributions centered on their literature values of 2.691~MeV and 
4.301~MeV, with $\sigma= 5$~keV and 12~keV, respectively, 
reflecting the experimental uncertainties~\cite{TILLEY2004}.

\begin{figure*}[htb]
\centering
\includegraphics[width=\textwidth]{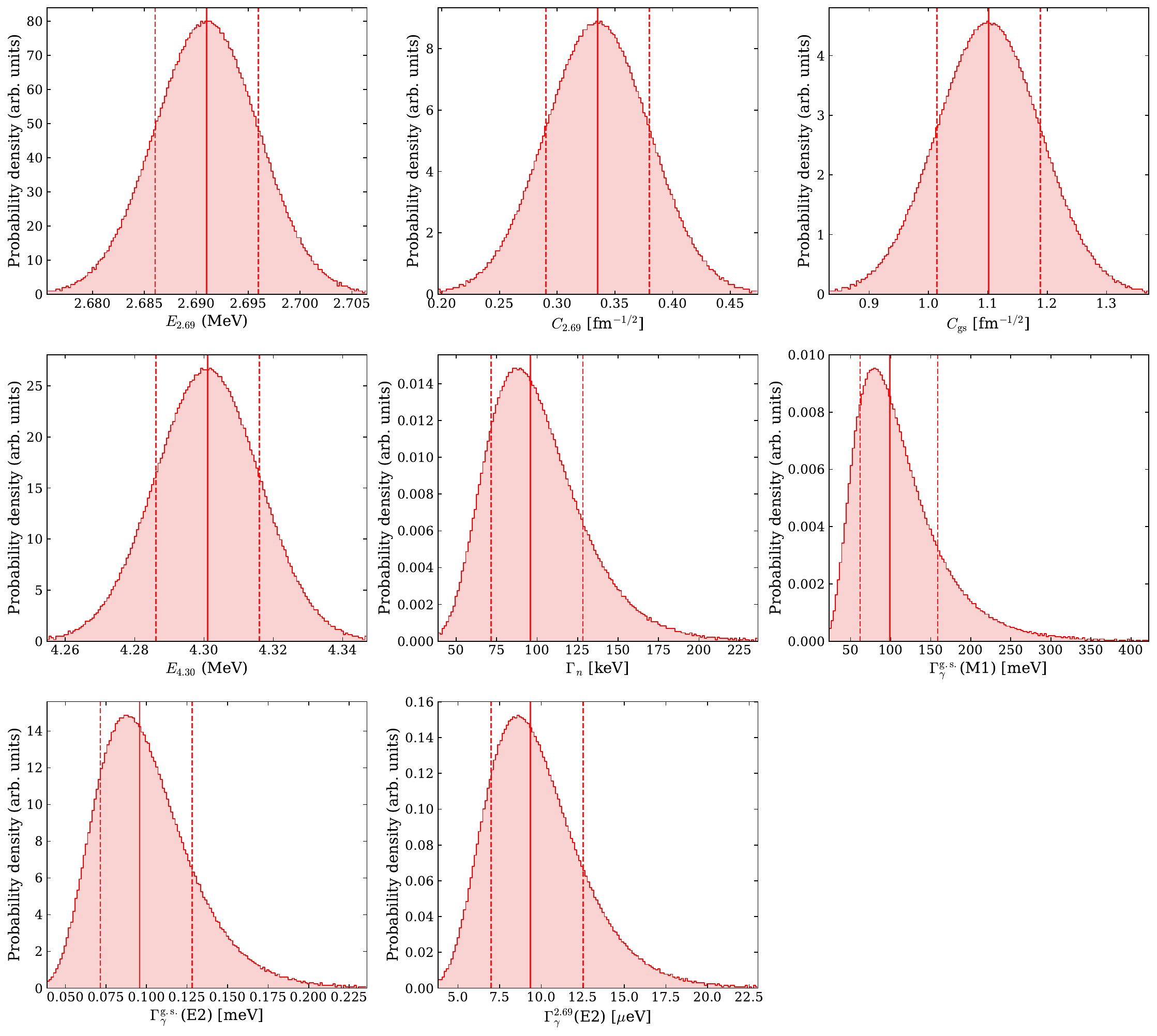}
\caption{\label{fig1} Probability density functions (PDFs) used for the Monte Carlo (MC) sampling of the $R$-matrix input parameters for the $^8$Li(n,$\gamma$)$^9$Li reaction. The channel radius is kept fixed at $r_c=4.5$~fm for the MC sampling. The vertical dashed lines indicate the 16\% and 84\% quantiles and the solid line represent the 50\% quantile of the
sampled distributions. See text for details.}
\end{figure*}
\begin{table}[htb]
\caption{Summary of bound and resonance state parameters of $^9$Li used in the present $R$-matrix calculation for the $^8$Li(n,$\gamma$)$^9$Li reaction. The ANC ($C$) values (in fm$^{-1/2}$) for the $^9$Li bound states are taken from the literature, as indicated by the references. For the $5/2^{-}$ resonant state at $E_x = 4.30$~MeV, the partial neutron width ($\Gamma_n$ in keV) and $\gamma$-decay widths ($\Gamma_\gamma$ in meV)  are also listed. See text for details.
}
\label{tab:9Li_parameters}
\begin{ruledtabular}
\begin{tabular}{ccccc}
$E_x$ & $J^\pi$ & $C$/$\Gamma_n$ & $\Gamma_\gamma^{E_f}$ \\
(MeV) &  & (fm$^{-1/2}$/keV) & (meV) \\
\hline
g.s. & $3/2^-$ & 1.15(14)~\cite{GUO2005}  & -- \\
     &         & 1.026, 0.995~\cite{McCracken2021}\\
     &         & 1.12(11)~\cite{Li2005,LIU06} & \\
     &         & 0.96(8)~\cite{Guimaraes2007} &\\
     &         & 1.12~\cite{HUANG2010} & \\
     &         & 0.98, 1.19, 1.10~\cite{Tengborn2011} &\\
     &         & 1.23(6), 1.180(15)~\cite{Sargsyan2023} &\\    
     &         & 1.181(13)~\cite{Nollett2011} &\\ 
     &         & 1.088~\cite{Timofeyuk2013} &\\ [3ex]
$2.691(5)$ & $1/2^-$ & 0.40~\cite{HUANG2010}  & -- \\
     &         & 0.298~\cite{Tengborn2011} & \\
     &         & 0.308(7)~\cite{Nollett2011} & \\ 
     &         & 0.33~\cite{Timofeyuk2013}  & \\ [3ex]
$4.301(12)$ &$5/2^-$  & 100(30)~\cite{Wuosmaa2005} & $35,55,111$~($M1$)$^{\text{g.s.}}$~\cite{Ma2012}  \\
&      &79.0, 77.8~\cite{Ma2012}    &$0.1$~($E2$)$^{\text{g.s.}}$~\cite{Caprio2022} \\
&      &88(25)~\cite{TILLEY2004}  &$9.77\times 10^{-3}$~($E2$)$^{\text{2.69}}$  \\
\end{tabular}
\end{ruledtabular}
\end{table}

\begin{table}
\caption{Summary of the Monte Carlo sampled $R$-matrix input parameters for the
$^8$Li(n,$\gamma$)$^9$Li reaction. Gaussian distributions $\mathcal{N}(\mu,\sigma)$
are used for sampling the $^9$Li level energies and ANCs, log-normal distributions
$\ln\mathcal{N}(\mu,\sigma)$, parameterized by their mean $\mu$ and standard deviation
$\sigma$, are used for the partial widths. Quoted results are the median values, defined by the 50\% quantile,
together with the lower and upper MC uncertainties defined by the
16\% and 84\% quantiles of the sampled distributions. See text for details.}
\label{tab:mc_params}
\begin{ruledtabular}
\begin{tabular}{lcc}
Parameter [Unit] & MC Sampling PDF & Median$_{-16\%}^{+84\%}$ \\
\hline
$E_{2.69}$ [MeV]                             & $\mathcal{N}(2.691,\;0.005)$  & $2.691^{+0.005}_{-0.005}$ \\[3pt]
$C_{2.69}$ [fm$^{-1/2}$]                    & $\mathcal{N}(0.335,\;0.045)$  & $0.335^{+0.045}_{-0.045}$ \\[3pt]
$C_{\mathrm{gs}}$ [fm$^{-1/2}$]             & $\mathcal{N}(1.102,\;0.088)$  & $1.102^{+0.088}_{-0.088}$ \\[3pt]
$E_{4.30}$ [MeV]                             & $\mathcal{N}(4.301,\;0.012)$  & $4.301^{+0.012}_{-0.012}$ \\[3pt]
$\Gamma_{n}$ [keV]                           & $\ln\mathcal{N}(100,\;30)$    & $95.8^{+32.4}_{-24.3}$    \\[3pt]
$\Gamma_{\gamma}^{\mathrm{g.s.}}(M1)$ [meV] & $\ln\mathcal{N}(111,\;55.5)$  & $99.3^{+59.5}_{-37.2}$    \\[3pt]
$\Gamma_{\gamma}^{\mathrm{g.s.}}(E2)$ [meV] & $\ln\mathcal{N}(0.10,\;0.03)$ & $0.096^{+0.032}_{-0.024}$ \\[3pt]
$\Gamma_{\gamma}^{2.69}(E2)$ [$\mu$eV]      & $\ln\mathcal{N}(9.77,\;2.93)$ & $9.36^{+3.17}_{-2.37}$    \\
\end{tabular}
\end{ruledtabular}
\end{table}

Following Longland \textit{et al.}~\cite{Longland2010}, the partial widths
$\Gamma_n$ and $\Gamma_\gamma$ are assigned log-normal distributions. The neutron width $\Gamma_n$ of the $5/2^-$ resonance is informed by the 
measurement of Ref.~\cite{Wuosmaa2005}, which reports $\Gamma_n = 100(30)$~keV 
and serves as the mean value of the log-normal sampling distribution 
$\ln\mathcal{N}(100,\,30)$~keV. In contrast, no experimental data exist for 
the $\gamma$-decay widths $\Gamma_\gamma$. Shell-model calculations predict 
$M1$ electromagnetic decay widths for the $5/2^- \rightarrow 3/2^-$ 
ground-state transition ranging from 35 to 111~meV~\cite{Ma2012}, as 
summarized in Table~\ref{tab:9Li_parameters}. The width 
$\Gamma_\gamma^{\rm g.s.}(M1)$ is sampled from a log-normal distribution 
$\ln\mathcal{N}(111,\,55.5)$~meV, centered on the largest shell-model 
prediction~\cite{Ma2012}. The $E2$ decay width for the 
$5/2^- \rightarrow 3/2^-$ ground-state transition is calculated using the 
\textit{ab initio} estimate of $B(E2)$ from Ref.~\cite{Caprio2022}, while the 
$E2$ width for the $5/2^- \rightarrow 1/2^-$ transition to the 
$E_x=2.69$~MeV state is derived from the Weisskopf 
estimate~\cite{iliadis2007}. The sampling distributions for all eight 
parameters are summarized in Table~\ref{tab:mc_params} and the plots are shown in Fig.~\ref{fig1}.

\section{R-matrix model uncertainty}~\label{sec:model_uncertainty}
\begin{figure}[htb]
\centering
\includegraphics[width=0.5\textwidth]{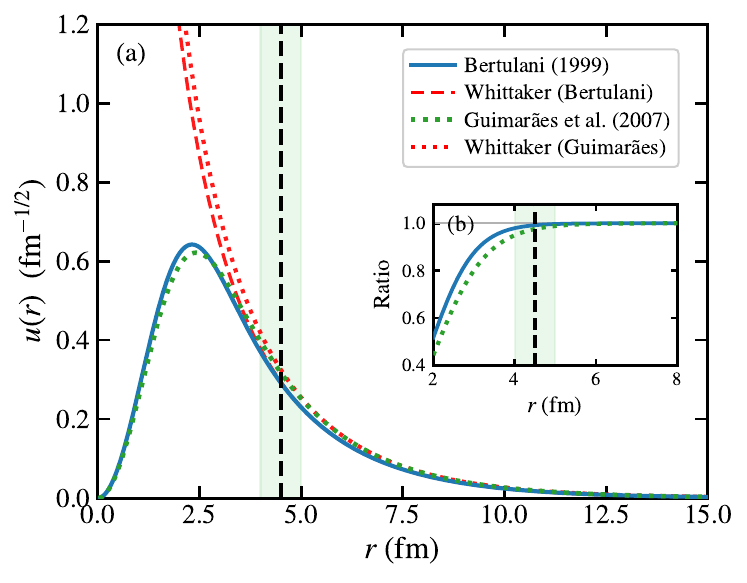}
\caption{ \label{fig:rc_window}
(a) Comparison of the normalized radial bound-state wave function $u(r)$ for the $^9$Li ground state (blue solid and green dotted lines) with the corresponding asymptotic forms (red dashed and red dotted lines, respectively) for the single-particle potentials of Bertulani~\cite{Bertulani1999} and Guimar{\~a}es et al.~\cite{Guimaraes2007}. (b) Ratio of each bound-state wave function to its asymptotic form, normalized to unity at large radius. The shaded band marks the adopted interval $r_c = 4.0–5.0$~fm and the vertical dashed line the central value $r_c = 4.5$~fm.
}
\end{figure}
\begin{figure}[htb]
\centering
\includegraphics[width=0.5\textwidth]{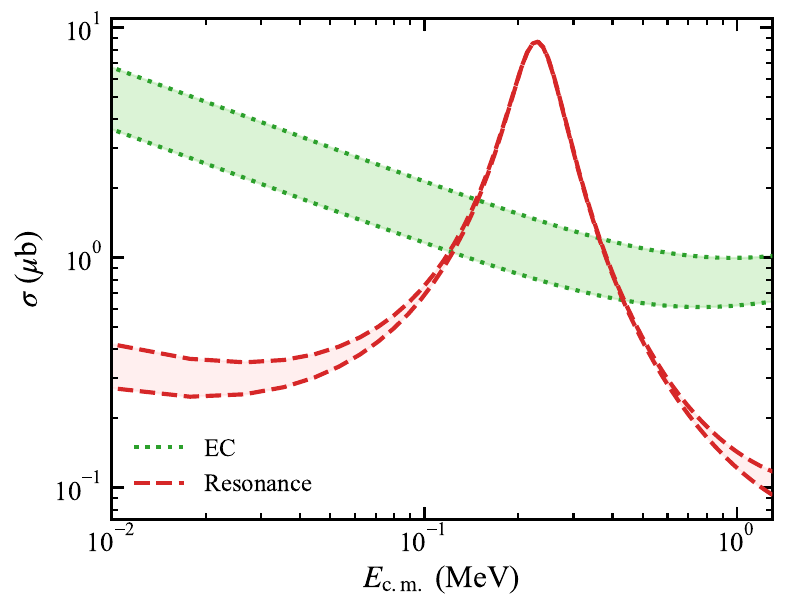}
\caption{ \label{fig:rc_sensitivity}
Sensitivity of the external-capture (EC) and resonant-capture
components of the $^8$Li$(n,\gamma)^9$Li cross section to the channel
radius, $r_c$. The green and red shaded bands show the variations of
the EC and resonant-capture contributions, respectively, over
$r_c=4.0$--$5.0$~fm. For the EC component, the lower and upper dotted
curves correspond to $r_c=5.0$ and $4.0$~fm, respectively. The EC contribution exhibits the
stronger sensitivity to the $r_c$ variation.
}
\end{figure}
As discussed in section~\ref{sec:R-matrix}, the channel radius ($r_c$) is treated as a model parameter of the $R$-matrix model,
and the resulting sensitivity of the calculated cross section is
quoted as a model uncertainty,  following
Ref.~\cite{deBoer2014}.  
We note an important difference from fitted
analysis in Ref.~\cite{deBoer2014}, the acceptable range of $r_c$ was
defined by the goodness of the fit ($\chi^2$ minimization), and the fitted background-pole
parameters readjusted with $r_c$ so as to largely compensate the radius
dependence of the external capture.  
In the present unfitted calculation, the adopted range $r_c = 4.0-5.0$~fm is
instead motivated not only by the standard prescription of $r_c=1.4(A_1^{1/3}+A_2^{1/3})$ and the $r_c$ value adopted in previous $R$-matrix analysis of $^7$Li+$n$ system as discussed in section~\ref{sec:R-matrix}, but also the asymptotic behaviour of the $^9$Li ground state wave function.

To place the adopted $r_c$ range on a physical basis, the radial distance at which the $^9$Li ground-state
single-particle wave function reaches its asymptotic form is examined. The
$^9$Li$(3/2^-)$ ground state is described as a $p_{3/2}$ neutron
coupled to the $^8$Li$(2^+)$ core using Woods-Saxon geometries
adopted in previous potential model
calculations~\cite{Bertulani1999,Guimaraes2007}, with the
central potential depth in each case adjusted to reproduce the
corresponding one-neutron separation energy.

The normalized radial wave functions are compared with the neutron $p$-wave
Whittaker asymptotic form as shown in Fig.~\ref{fig:rc_window}(a). This comparison provides a direct indication
of the radial distance beyond which the bound-state wave function is
determined predominantly by its asymptotic tail.
For the geometry of Bertulani~\cite{Bertulani1999},
the wave function reaches its asymptotic Whittaker form to within
5\%, 2\%, and 1\% for radii larger than
$r_c=3.5$, $4.0$, and $4.4$~fm, respectively. For the more diffuse
Guimar{\~a}es et al.~\cite{Guimaraes2007} geometry, the corresponding
radii are shifted outward to approximately $r_c=4.0$, $4.6$, and
$5.1$~fm, as shown in Fig.~\ref{fig:rc_window}(b). 
The shaded
region marks the adopted interval $r_c=4.0-5.0$~fm and the vertical
dashed line indicates the central value $r_c=4.5$~fm used for the
Monte Carlo uncertainty propagation. Over this interval the
bound-state overlap is predominantly asymptotic for both
potentials. 

In the present calculation, the non-resonant DC component corresponds
to the pure external-capture (EC) term in the hard-sphere formulation~\cite{AZURE}. In this formulation, the external scattering
wave function entering the capture matrix element contains the
hard-sphere phase shift, as discussed by Azuma et al.~\cite{AZURE}.
Consequently, the EC amplitude depends on the channel radius not only
through the radial integral, but also through the
hard-sphere phase shift of the incoming scattering wave. No fitted
internal non-resonant capture amplitude is included in the present
calculation.
Fig.~\ref{fig:rc_sensitivity} shows the sensitivity of the
EC and resonant-capture components to the channel
radius over the interval $r_c=4.0-5.0$~fm. The calculations are
performed using the median values of all $R$-matrix input
parameters listed in Table~\ref{tab:mc_params}. The EC contribution
exhibits a clear $r_c$ dependence, the EC cross section obtained with
$r_c=4.0$~fm is larger than that obtained with $r_c=5.0$~fm, differing
by a factor of $\approx 1.5-2.0$. By contrast, the resonant-capture
component is relatively insensitive to change in $r_c$. Thus, the channel-radius
model uncertainty in the total capture cross section is dominated by
the sensitivity of the EC component. 

\section{Results} \label{sec:Results}

\begin{figure}[htb]
\centering
\includegraphics[width=0.5\textwidth]{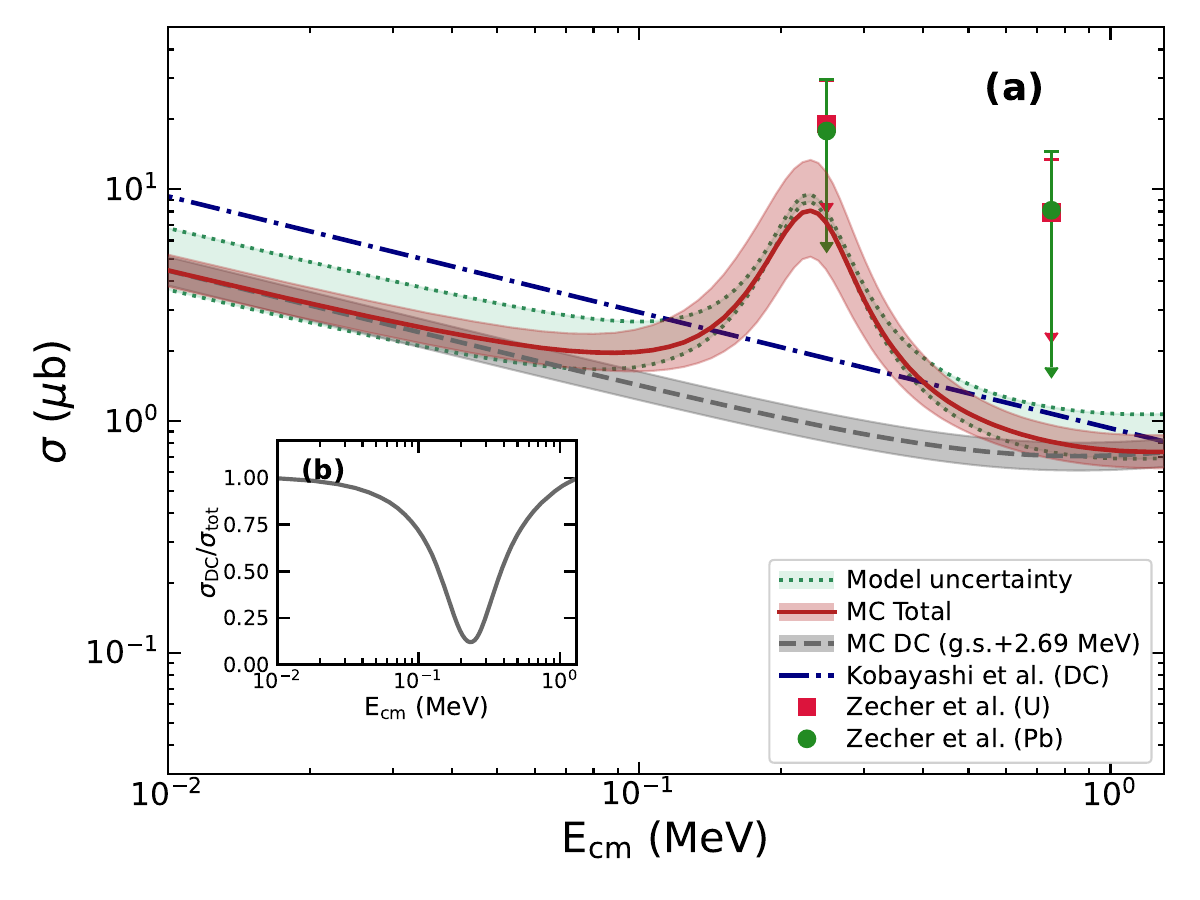}
\caption{ \label{fig:xs}
Radiative neutron-capture cross section ($\sigma$) for the $^{8}$Li(n,$\gamma$)$^{9}$Li reaction from the present $R$-matrix calculations. Panel~(a) shows the direct-capture (DC) contribution and the total cross section, including both DC and resonance, as a function of the center-of-mass energy ($E_{\mathrm{c.m.}}$). The experimental upper limits from Zecher et al.~\cite{Zecher1998}, obtained using U and Pb targets, and the DC upper limit of Kobayashi et al.~\cite{Kobayashi2003} (blue dotted line) are also shown for comparison. The green shaded band represents the $R$-matrix model uncertainty
associated with the channel-radius variation over
$r_c=4.0-5.0$~fm. The red and gray shaded bands represent the MC uncertainty bands,
bounded by the 16\% and 84\% quantiles of the 10000-sample ensemble
for the total and DC cross sections, respectively.
Panel~(b) shows the ratio of the median DC cross section to the median total cross section from MC sampling.
}
\end{figure}

\begin{figure}[htb]
\centering
\includegraphics[width=0.5\textwidth]{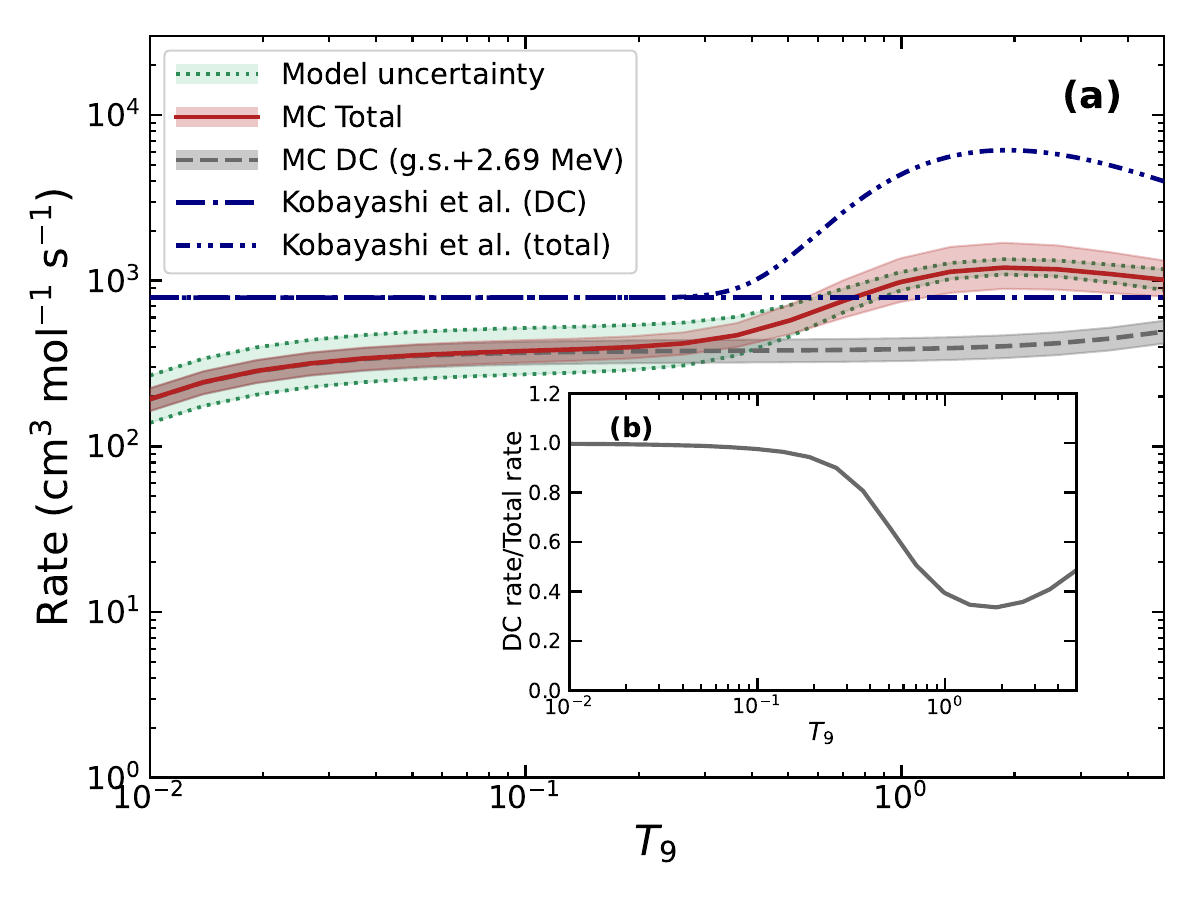}
\caption{\label{fig:rate}Reaction rate for the $^{8}$Li(n,$\gamma$)$^{9}$Li reaction from the present $R$-matrix calculations. Panel~(a) shows the DC contribution and the total reaction rate, including both DC and resonance, as a function of temperature ($T_9$). The red and gray shaded bands represent the MC uncertainty bands,
bounded by the 16\% and 84\% quantiles of the 10000-sample ensemble
for the total and DC rates, respectively. The green shaded band represents the
separate $R$-matrix model uncertainty associated with the
channel-radius variation over $r_c=4.0$--$5.0$~fm. The upper limit of the total reaction rate and the DC rate reported by Kobayashi et al.~\cite{Kobayashi2003} are shown by the blue dash-dotted and dotted lines, respectively.  Panel~(b) shows the ratio of the median DC reaction rate to the median total reaction rate. See text for details.}
\end{figure}

\begin{figure}[htb]
\centering
\includegraphics[width=0.5\textwidth]{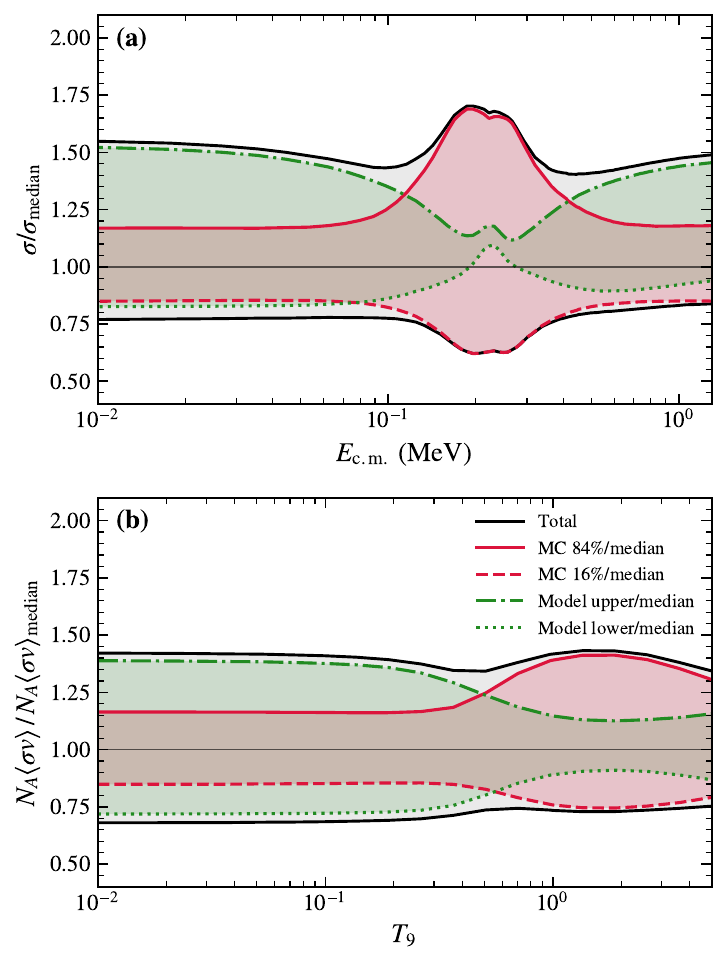}
\caption{\label{fig:uncertainty}Fractional uncertainties of the $^8$Li$(n,\gamma)^9$Li
(a) total capture cross section and (b) thermonuclear reaction rate,
shown as ratios to the MC median. The red band represents the MC
uncertainty band bounded by the 16\% and 84\% quantiles of the
10000-sample ensemble. The green band represents the 
$R$-matrix model uncertainty, bounded by calculations performed at
$r_c=4.0$ and $5.0$~fm with all other input parameters fixed at their
median values listed in Table~\ref{tab:mc_params}. The black
curves show an effective total uncertainty obtained by
adding the MC and model uncertainties in quadrature. }
\end{figure}

The radiative neutron capture cross section and the corresponding thermonuclear reaction rate for the $^8$Li(n,$\gamma$)$^9$Li reaction from the present $R$-matrix calculations are shown in Fig.~\ref{fig:xs} and Fig.~\ref{fig:rate}, respectively.
The MC sampled median total capture cross section (MC Total) consisting of both the DC and resonant contributions as a function of the center-of-mass energies ($E_\textrm{c.m.}$) is shown with a red solid line in Fig.~\ref{fig:xs}(a).
The median DC (MC DC) component due to the non-resonant capture to the g.s. and to the 2.69 MeV excited state is plotted as a black dashed line.
The red and gray shaded bands represent the MC uncertainty bands, bounded by the 16\% and 84\% quantiles of the 10000-sample ensemble
for the total and DC cross sections, respectively.
The green shaded band in
Fig.~\ref{fig:xs}(a) shows the separate $R$-matrix model uncertainty
associated with the channel-radius variation over
$r_c=4.0-5.0$~fm, calculated using the median values of all
MC-sampled input parameters listed in Table~\ref{tab:mc_params}. This model uncertainty is energy dependent and is shown in
Fig.~\ref{fig:uncertainty}(a) as a ratio to the MC median total cross
section. 
It should be noted that the MC median and its 68\% uncertainty interval
are obtained from an ensemble of 10000 sampled $R$-matrix calculations
performed with the channel radius fixed at the central value
$r_c=4.5$~fm. In contrast, the $R$-matrix model uncertainty associated
with the channel radius is evaluated separately by varying
$r_c=4.0$--$5.0$~fm while keeping all other input parameters fixed at
their MC sampled median values listed in Table~\ref{tab:mc_params}.


In the low-energy EC-dominated region, the model uncertainty band is
wider than the MC uncertainty band, spanning approximately $0.83$--$1.52$ of the MC
median at $E_{\rm c.m.}\approx 0.01$~MeV and about $0.86$--$1.35$ near
$E_{\rm c.m.}\approx 0.1$~MeV. Near the resonant peak, the
channel-radius dependence is strongly reduced, around
$E_{\rm c.m.}\approx 0.24$--$0.25$~MeV, the model band contracts to
approximately $1.05$--$1.14$ of the MC median, while the MC band expands
substantially because of the sampled resonance parameters. At higher
energies, where the EC component again gives the dominant
contribution, the model band widens again, reaching about
$0.90$--$1.46$ of the MC median over
$E_{\rm c.m.}=0.75$--$1.32$~MeV. Thus, the $R$-matrix model
uncertainty is important in the EC-dominated regions, whereas
the resonant peak is governed primarily by the MC uncertainty of the
sampled resonance parameters.
An effective
total uncertainty is constructed by adding the MC and model uncertainties
in quadrature as shown by the black lines in Fig.~\ref{fig:uncertainty}. 

Experimental data from Zecher~\textit{et al.}~\cite{Zecher1998} are included for comparison in Fig.~\ref{fig:xs}(a), where measurements with uranium (U) and lead (Pb) targets are depicted by red squares and green circles, respectively.
At $E_\textrm{c.m.} = 0.25$ and 0.75~MeV, the present total capture cross section is within the upper limits reported by Zecher~\textit{et al.}~\cite{Zecher1998}.
It should be noted that the data by Zecher~\textit{et al.}~\cite{Zecher1998} are averaged over broad energy intervals of $E_\text{c.m.}=0–0.5$~MeV and 0.5–1.0~MeV. 
The total capture cross section in Fig.~\ref{fig:xs}(a) is dominated by the non-resonant DC component over most of the energy range, except in the region $E_{\mathrm{c.m.}}=0.1-0.5$~MeV, where the $5/2^-$ resonance dominates. This is evident from the inset in Fig.~\ref{fig:xs}(b), which shows the ratio of the median DC cross section to the median total cross section from the MC sampling.

\begin{figure*}
    \centering
    \includegraphics[width=\textwidth]{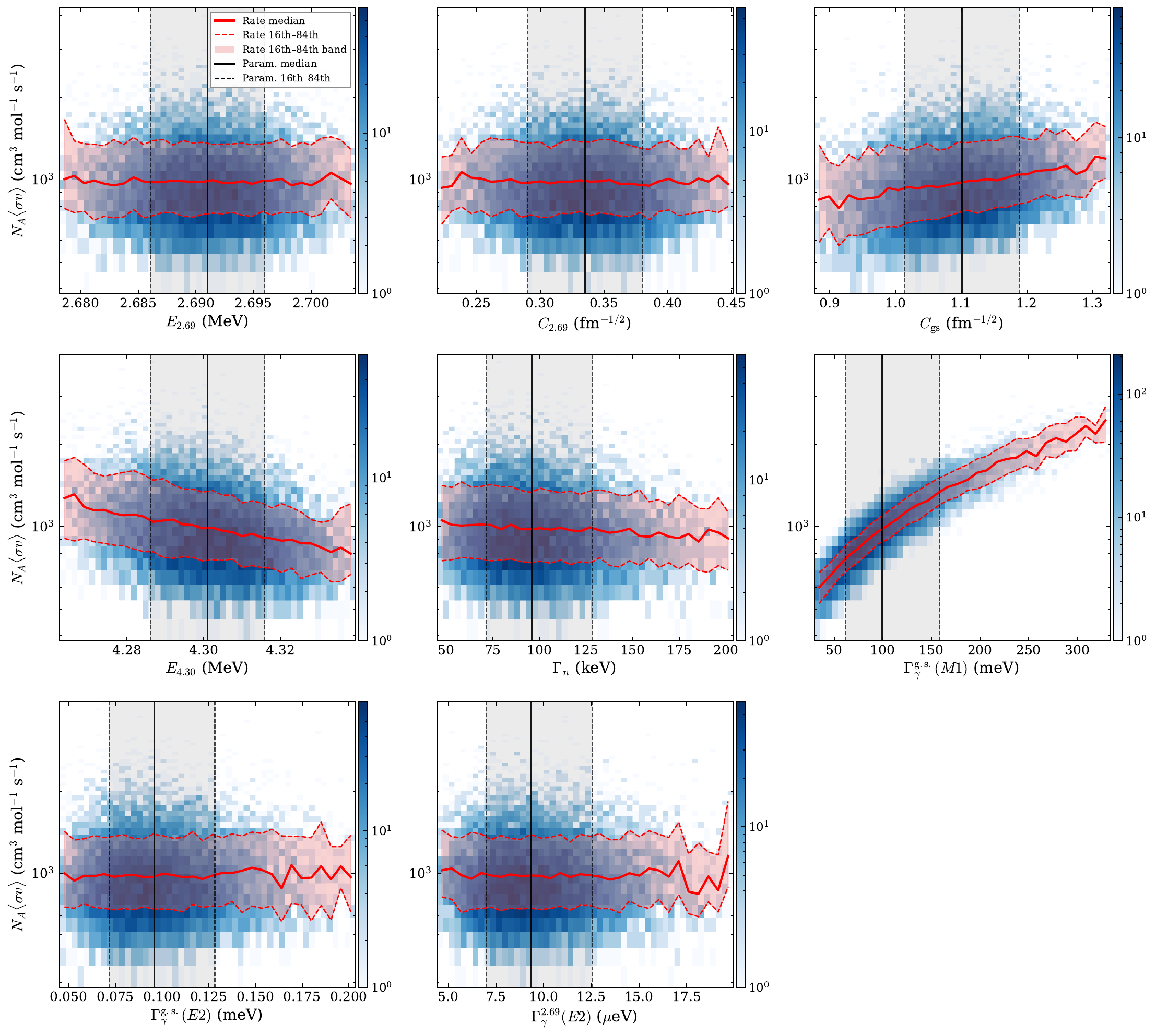}
    \caption{Thermonuclear reaction rate $N_A\langle\sigma v\rangle$ at 
    $T_9 = 1.0$ as a function of each MC sampled $R$-matrix 
    input parameter. Each panel shows the distribution of 10000 MC samples 
    (blue density), with the running median rate (solid red line) and 
    the 16\%-84\% MC uncertainty interval (shaded red band, bounded by dashed red lines) 
    computed in bins along the parameter axis. Vertical lines indicate the 
    median (solid black) and 16th-84th percentile interval (dashed black) 
    of each parameter's sampling distribution. The reaction rate at 
    $T_9 = 1.0$ is most sensitive to $\Gamma_\gamma^{\rm g.s.}(M1)$, 
    which shows a strong positive correlation with the rate, while all 
    other parameters are relatively flat. See text for details.}
    \label{fig:rate_sensitivity}
\end{figure*}

\begin{table*}[htbp]
\centering
\caption{Direct-capture ($\sigma_{\rm DC}$) and total capture  ($\sigma_{\rm total}$) (DC+Resonance)
cross sections for the $^{8}{\rm Li}(n,\gamma)^{9}{\rm Li}$ reaction from the
present work and previous studies. The uncertainties in the present work represent the effective total uncertainties, obtained by adding the Monte Carlo propagated uncertainties and the $R$-matrix model uncertainty in quadrature. Cross sections quoted as upper limits
are preceded by ``$<$'', and values preceded by ``$\sim$'' are estimated from the
published figures of the cited references. A dash indicates that the corresponding
value is not reported.}
\label{tab:xsec_literature}
\begin{ruledtabular}
\begin{tabular}{c c c l c}
$E_{\rm c.m.}$ (MeV) 
& $\sigma_{\rm DC}$ ($\mu$b) 
& $\sigma_{\rm total}$ ($\mu$b) 
& Reference 
& Year \\
\hline
\multirow{3}{*}{$\sim10^{-8}$}
& $3.79\times10^{4}$ 
& -- 
& Descouvemont~\cite{Descouvemont1993} 
& 1993 \\
& $3.4\times10^{4}$ 
& -- 
& Rauscher~\cite{Rauscher1994} 
& 1994 \\
& -- 
& $(3.82$--$7.45)\times10^{4}$ 
& Dubovichenko \textit{et al.}~\cite{Dubovichenko2023} 
& 2023 \\
\hline
\multirow{4}{*}{0.020--0.025}
& $38$      
& -- 
& Descouvemont~\cite{Descouvemont1993} 
& 1993 \\
& $<5.88$ 
& -- 
& Kobayashi~\cite{Kobayashi2003}, Mohr~\cite{Mohr2003} 
& 2003 \\
& $\sim 50$    
& $\sim 50$ 
& Ma \textit{et al.}~\cite{Ma2012}\footnotemark[1]  
& 2012 \\
& --   
& $\sim 95$ 
& McCracken \textit{et al.}~\cite{McCracken2021} 
& 2021 \\
& $2.82^{+1.50}_{-0.65}$   
& $2.91^{+1.50}_{-0.65}$
& \textbf{Present work}
& 2026 \\
\hline
\multirow{9}{*}{0.25}
& $<19.1\pm10.4$  
& --
& Zecher \textit{et al.}~\cite{Zecher1998} (U) 
& 1998 \\
&  $<17.8\pm11.9$ 
&--
& Zecher \textit{et al.}~\cite{Zecher1998} (Pb) 
& 1998 \\
& $\sim 5$  
& -- 
& Bertulani~\cite{Bertulani1999}\footnotemark[1] 
& 1999 \\
& $<1.86$ 
& -- 
& Kobayashi \textit{et al.}~\cite{Kobayashi2003}\footnotemark[1] 
& 2003 \\
& $\sim 8$  
& -- 
& Banerjee \textit{et al.}~\cite{Banerjee2008}\footnotemark[1] 
& 2008 \\
& $\sim 8$ 
& -- 
& Huang \textit{et al.}~\cite{HUANG2010}\footnotemark[1] 
& 2010 \\
& $\sim 9$ 
& $\sim16$--30 
& Ma \textit{et al.}~\cite{Ma2012}\footnotemark[1] 
& 2012 \\
& --   
& $\sim 40$ 
& McCracken \textit{et al.}~\cite{McCracken2021} 
& 2021 \\
& 15.23 
& 16.57 
& Dong \textit{et al.}~\cite{Dong2023Erratum} 
& 2023 \\
& $0.94^{+0.48}_{-0.21}$ 
& $7.22^{+4.80}_{-2.70}$ 
& \textbf{Present work}
& 2026 \\
\hline
\multirow{9}{*}{0.75}
&  $< 7.9\pm5.5$ 
&--
& Zecher \textit{et al.}~\cite{Zecher1998} (U) 
& 1998 \\
&  $<8.1\pm6.4$ 
& --
& Zecher \textit{et al.}~\cite{Zecher1998} (Pb) 
& 1998 \\
& $\sim 3$  
& -- 
& Bertulani~\cite{Bertulani1999}\footnotemark[1] 
& 1999 \\
& $<1.07$ 
& -- 
& Kobayashi \textit{et al.}~\cite{Kobayashi2003}\footnotemark[1] 
& 2003 \\
& $< 6$  
& -- 
& Banerjee \textit{et al.}~\cite{Banerjee2008}\footnotemark[1] 
& 2008 \\
& $\sim 3$  
& -- 
& Huang \textit{et al.}~\cite{HUANG2010}\footnotemark[1] 
& 2010 \\
& $\sim 6$ 
& $\sim 6$ 
& Ma \textit{et al.}~\cite{Ma2012}\footnotemark[1] 
& 2012 \\
& --
& $\sim 10$ 
& McCracken \textit{et al.}~\cite{McCracken2021} 
& 2021 \\
& 5.24 
& 6.94 
& Dong \textit{et al.}~\cite{Dong2023Erratum} 
& 2023 \\
& $0.71^{+0.32}_{-0.13}$ 
& $0.81^{+0.36}_{-0.14}$ 
& \textbf{Present work }
& 2026 \\
\end{tabular}
\end{ruledtabular}
\footnotetext[1]{Value read from the published figure; not tabulated in the
cited reference.}
\end{table*}

A comparison of the total $^{8}$Li$(n,\gamma)^{9}$Li capture cross
section, including both the DC and $5/2^-$ resonant contributions,
and the DC component alone from the present $R$-matrix analysis with
available literature values is presented in
Table~\ref{tab:xsec_literature}. 
The present total capture cross
sections at $E_{\rm c.m.}=0.25$ and $0.75$~MeV are
\[
\sigma_{\rm tot}(0.25~{\rm MeV})
=
7.22^{+4.80}_{-2.70}({\rm total})~\mu{\rm b},
\]
and
\[
\sigma_{\rm tot}(0.75~{\rm MeV})
=
0.81^{+0.36}_{-0.14}({\rm total})~\mu{\rm b},
\]
respectively. The corresponding DC cross sections are
\[
\sigma_{\rm DC}(0.25~{\rm MeV})
=
0.94^{+0.48}_{-0.21}({\rm total})~\mu{\rm b},
\]
and
\[
\sigma_{\rm DC}(0.75~{\rm MeV})
=
0.71^{+0.32}_{-0.13}({\rm total})~\mu{\rm b}.
\]
Here, the quoted total uncertainties are effective uncertainties
obtained by adding the MC and channel-radius model uncertainties in
quadrature. The MC uncertainty is defined by the 16\% and 84\%
quantiles of the 10000-sample ensemble at fixed $r_c=4.5$~fm, while
the model uncertainty is obtained by varying $r_c=4.0$--$5.0$~fm with
all other input parameters fixed at their median values listed in
Table~\ref{tab:mc_params}.

The present DC cross sections are significantly lower than those predicted by potential model~\cite{Bertulani1999,Banerjee2008,HUANG2010,Dubovichenko2016}, shell model~\cite{Mao1991,Ma2012}, \textit{ab initio} NCSMC~\cite{McCracken2021} calculations and the recent GSM-CC results~\cite{Dong2023Erratum}. However, the present DC cross section is within the $2\sigma$ uncertainty limit obtained by the measurement of Kobayashi et al.~\cite{Kobayashi2003}.
A probable reason for this could be that the treatment of external capture in the $R$-matrix calculations by \texttt{AZURE2} differs from that in the potential model approaches used in the previous studies. In \texttt{AZURE2}, the DC component includes only the hard-sphere (external) contribution, while potential model calculations also incorporate additional internal nuclear contributions. The differences from shell model results can be attributed to the choice of spectroscopic factors used.

The $^8$Li$(n,\gamma)^9$Li reaction rate derived from the
cross sections shown in Fig.~\ref{fig:xs}, over the temperature
range $T_9=0.01$--5, is presented in Fig.~\ref{fig:rate}. The median
total neutron-capture rate is shown by the red solid line in
Fig.~\ref{fig:rate}(a), while the DC contribution is shown by the
black dashed line. The red and gray shaded bands represent the
central 68\% uncertainty intervals from the MC uncertainty propagation
for the total and DC reaction rates, respectively. The green shaded
band shows the separate $R$-matrix model uncertainty associated with
the channel-radius variation over $r_c=4.0$--$5.0$~fm, evaluated with
all MC-sampled input parameters fixed at their median values. The
ratio of the median DC rate to the median total capture rate is shown
in Fig.~\ref{fig:rate}(b).

From Fig.~\ref{fig:rate}(b), it is evident that the DC component provides
the dominant contribution to the neutron-capture rate at low
temperatures, $T_9=0.01$--0.4, whereas at higher temperatures,
$T_9=0.5$--5, resonant capture through the $5/2^-$ state becomes
dominant. 
The DC reaction rate at $T_9=1$, obtained from the present
$R$-matrix calculation, is
\[
N_A\langle\sigma v\rangle_{\rm DC}
=
386.1^{+161.2}_{-122.3}({\rm total})
~{\rm cm^3~mol^{-1}~s^{-1}} .
\]
This DC rate is
significantly lower than previous rates reported in the literature
(Table~\ref{tab:9Li rates}) and remains within the upper limit
suggested by Kobayashi et al.~\cite{Kobayashi2003}.
The total capture rate at $T_9=1$ is
\[
N_A\langle\sigma v\rangle_{\rm tot}
=
983.9^{+410.9}_{-261.5}({\rm total})
~{\rm cm^3~mol^{-1}~s^{-1}} .
\]
The quoted total uncertainty is the effective uncertainty obtained by
adding the MC uncertainty and the
channel-radius model uncertainty in quadrature.

The relative importance of the uncertainty sources is summarized in
Fig.~\ref{fig:uncertainty} (a) and (b), which show the MC uncertainty band, the
channel radius model uncertainty, and the effective total uncertainty
as ratios to the MC median for both the total capture cross section
and the thermonuclear reaction rate. The effective total uncertainty
is obtained by adding the asymmetric MC and model uncertainties in
quadrature.
From Fig.~\ref{fig:uncertainty}(b), at low temperatures ($T_9=0.01$--$0.4$) where the DC component controls the rate, the
channel-radius model uncertainty dominates over the MC uncertainty.
Above the crossover near $T_9\approx 0.5$, the resonant contribution
becomes dominant and the MC uncertainty associated with the sampled
resonance parameters becomes the leading contribution.

To further investigate the sensitivity of the total reaction rate to the
different $R$-matrix input parameters, the reaction rate obtained from the
10000 MC samples is plotted as a function of each sampled parameter at
$T_9=1$, as shown in Fig.~\ref{fig:rate_sensitivity}. For each parameter, the running median of
the reaction rate, together with the 16th-84th percentile interval, is shown
to identify possible correlations between the sampled input parameter and the
calculated rate. The results show that the total reaction rate is most
sensitive to the $\gamma$-decay width $\Gamma_{\gamma}^{\rm g.s.}(M1)$ for the
$5/2^- \rightarrow 3/2^-_{\rm g.s.}$ transition, which exhibits a clear
positive correlation with the rate. 
Since no experimental data exist for the $M1$ $\gamma$-decay width of the
$5/2^{-} \rightarrow 3/2^{-}$ transition, which is currently constrained only by
shell-model predictions spanning $35-111$~meV~\cite{Ma2012}, an experimental
determination or improved theoretical estimate of this width would most
effectively reduce the uncertainty in the $^{8}$Li$(n,\gamma)^{9}$Li reaction
rate.
The dependence on the other parameters such as level
energies, ANCs, neutron width, and $E2$ $\gamma$-decay widths is comparatively
weak over the sampled ranges.

\section{Conclusion} \label{sec:Conclusion}

We have performed a phenomenological $R$-matrix analysis of the
$^8$Li$(n,\gamma)^9$Li radiative-capture reaction over the
center-of-mass energy range $E_{\rm c.m.}=0.01$--1.32~MeV,
including both non-resonant direct capture to the ground and
2.69~MeV states and resonant capture through the $5/2^-$ state at
$E_x=4.30$~MeV. The calculations were carried out using
\texttt{AZURE2}~\cite{AZURE} interfaced with the \texttt{BRICK}~\cite{BRICK} package. The
uncertainties of the $R$-matrix input parameters were propagated
through 10000 Monte Carlo samples, yielding statistically meaningful MC uncertainty intervals,
defined by the 16\% and 84\% quantiles, for the calculated cross
sections and reaction rates. In addition, the sensitivity to the channel radius was
evaluated separately over $r_c=4.0$--$5.0$~fm and quoted as an
$R$-matrix model uncertainty. The effective total uncertainty is obtained by adding the Monte Carlo propagated uncertainty and the $R$-matrix model uncertainty in quadrature.

In the present calculation, the non-resonant DC component corresponds
to the pure external-capture contribution in the hard-sphere
formulation by Azuma et al.~\cite{AZURE}. The channel-radius model uncertainty is driven
mainly by the external-capture component and is most important in the
EC-dominated regions, whereas the resonant peak is governed primarily
by the Monte Carlo uncertainty of the sampled resonance parameters.
The total capture cross sections are consistent with the energy-averaged
upper limits reported by Zecher et al.~\cite{Zecher1998}.

The corresponding thermonuclear reaction rate is dominated by direct
capture at low temperatures, $T_9=0.01$--0.4, while resonant capture
through the $5/2^-$ state dominates at higher temperatures,
$T_9=0.5$--5. 
The present DC rate is within the upper limit suggested by
Kobayashi et al.~\cite{Kobayashi2003}, while many previous theoretical predictions exceed
this limit. The total capture rate from the present analysis is $983.9^{+410.9}_{-261.5}~\mathrm{cm^{3}\,mol^{-1}\,s^{-1}}$ at $T_9 = 1$.

The parameter-sensitivity analysis at $T_9=1$ identifies the $M1$
$\gamma$-decay width of the
$5/2^- \rightarrow 3/2^-_{\rm g.s.}$ transition as the most sensitive
input parameter for the total reaction rate. This width is currently
constrained only by shell-model predictions and lacks an experimental
determination. A direct experimental constraint, or an improved
theoretical estimate, of this $\gamma$-width would therefore be the
most effective way to reduce the uncertainty in the
$^8$Li$(n,\gamma)^9$Li reaction rate in the resonance-dominated
temperature region. In contrast, the uncertainty in the
DC-dominated regime is mainly limited by the $R$-matrix model
treatment of the non-resonant external-capture contribution.
Further experimental constraints on the $^8$Li+n system, including scattering data and improved resonance parameter information,
are required to reduce the uncertainties to better constrain the
$^8$Li$(n,\gamma)^9$Li reaction rate.\\\\

\noindent
{\large{\bf Acknowledgements}}\\
Sk Mustak Ali thanks Richard James deBoer of the University of Notre Dame, and Sayan Samanta and Ritankar Mitra of Bose Institute, for fruitful discussions. Rajkumar Santra acknowledges financial support vide Ref. No. ANRF/ARG/2025/003936/PS.

\bibliography{Ref}

\end{document}